\documentclass[10pt,aps,prl,twocolumn,showpacs,amssymb,floatfix]{revtex4-1}
\usepackage{amsmath,graphics,epsfig,mathrsfs}
\usepackage{bm}
\usepackage{epigraph}

\begin{document}
\title{Controlling the Spin Torque Efficiency with Ferroelectric Barriers}
\author{A. Useinov$^{1}$}
\author{A. Manchon$^{1}$}
\email{aurelien.manchon@kaust.edu.sa}
\affiliation{$^1$ King Abdullah University of Science and Technology (KAUST), Physical Science and Engineering Division, Thuwal 23955-6900, Saudi Arabia.}
\date{\today}

\begin{abstract}
Non-equilibrium spin-dependent transport in magnetic tunnel junctions comprising a ferroelectric barrier is theoretically investigated. The exact solutions of the free electron Schr\"odinger equation for electron tunneling in the presence of interfacial screening are obtained by combining Bessel and Airy functions. We demonstrate that the spin transfer torque efficiency, and more generally the bias-dependence of tunneling magneto- and electroresistance, can be controlled by switching the ferroelectric polarization of the barrier. This effect provides a supplementary way to electrically control the current-driven dynamic states of the magnetization and related magnetic noise in spin transfer devices.
\end{abstract}
\maketitle

The electrical control of magnetization in thin films is currently attracting intensive efforts due to its major potential for applications \cite{reviewChappert}. Current- and voltage-induced magnetization dynamics have been observed in a wide variety of magnetic heterostructures such as metallic and semiconducting spin-valves and domain walls \cite{Brataas,Ralph,chapter,pma}. Of most technological interest is the manipulation of magnetization in magnetic tunnel junctions (MTJs), trilayers composed of a tunnel barrier embedded between two ferromagnets, by means of either gate voltages or spin-polarized currents. In the first case, a large voltage pulse is applied across the barrier and charge reordering at the interface modifies the interfacial magnetic anisotropy \cite{pma}. In the second case, a spin-polarized current tunnels through the barrier and transfers its spin angular momentum to the local magnetization of the free magnetic layer \cite{Brataas,Ralph,chapter}. This last phenomenon, known as {\em spin transfer torque} \cite{slonc96}, enables the design of promising components such as on-chip tunable microwave generators \cite{houss}, magnetic memory cells \cite{ieee}, race-track memories \cite{racetrack}, and so on.\par

The most successful candidate to date that combines efficient electrical manipulation with large {\em tunneling magnetoresistance} (TMR) ratios are MTJs based on MgO barriers and transition metal electrodes \cite{mgo}. Significant progress have been achieved towards understanding the complex microscopic nature of the junction's interfaces \cite{bonell,velev} and establishing the characteristics of spin transfer torque \cite{Theo,asym}. Despite undeniable successes \cite{stt_mtj,spindiode}, the actual exploitation of spin transfer torque in devices is facing major hurdles, among which its large critical current threshold as well as current-driven magnetic instabilities. Serious efforts are being made towards improving the device performances and solutions such as engineering the junction structural asymmetries \cite{Oh} or designing the metallic electrodes stacking \cite{diff} have been proposed to ameliorate of the device operation through modifying the bias dependence of the spin torque. However, little has been done on the barrier materials itself and MgO is still considered as the best candidate for spin torque purposes.

\begin{figure}
\begin{tabular}{c}
\includegraphics[angle=270, width=6.2cm]{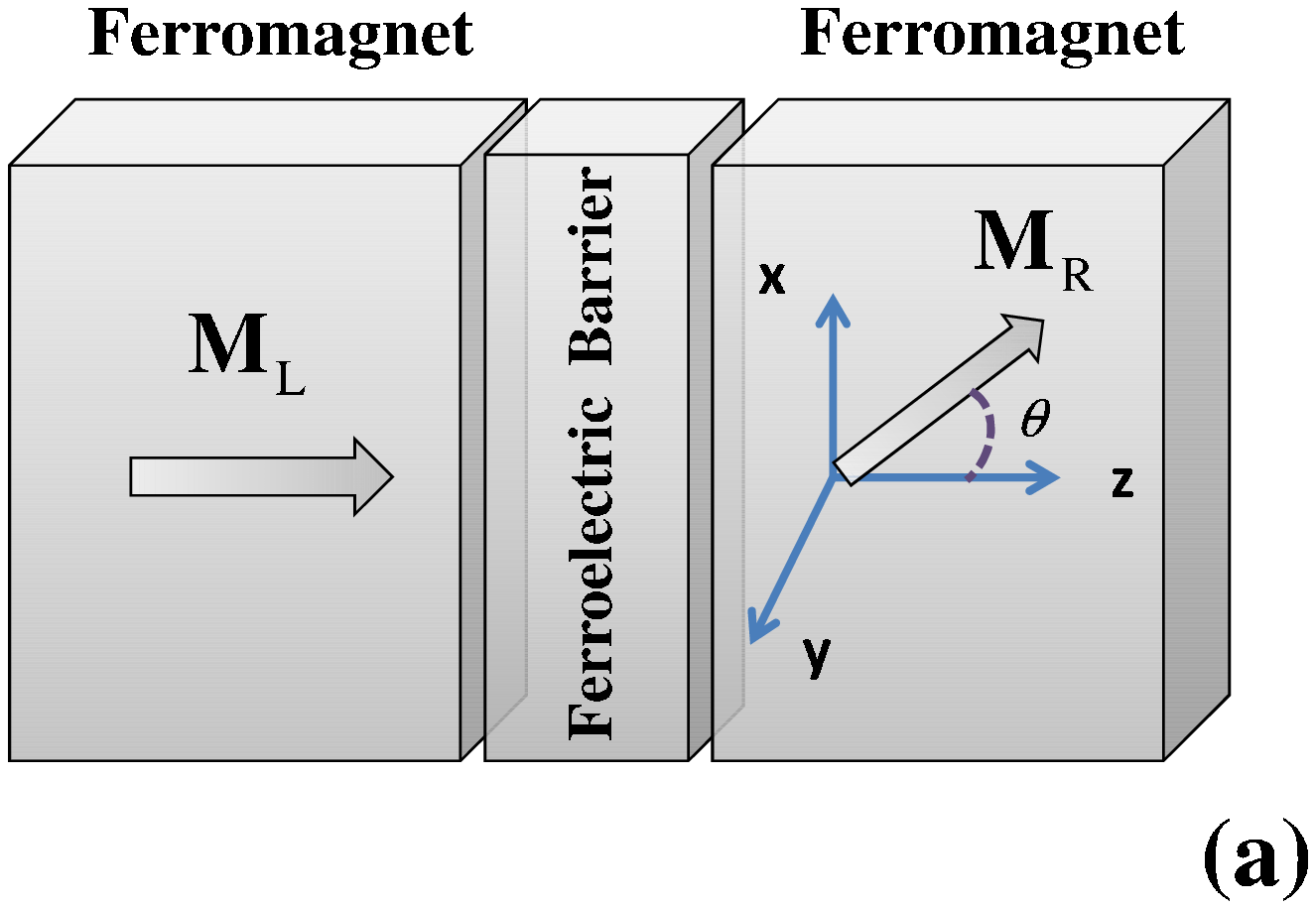}\\
\\
\includegraphics[width=7.6cm]{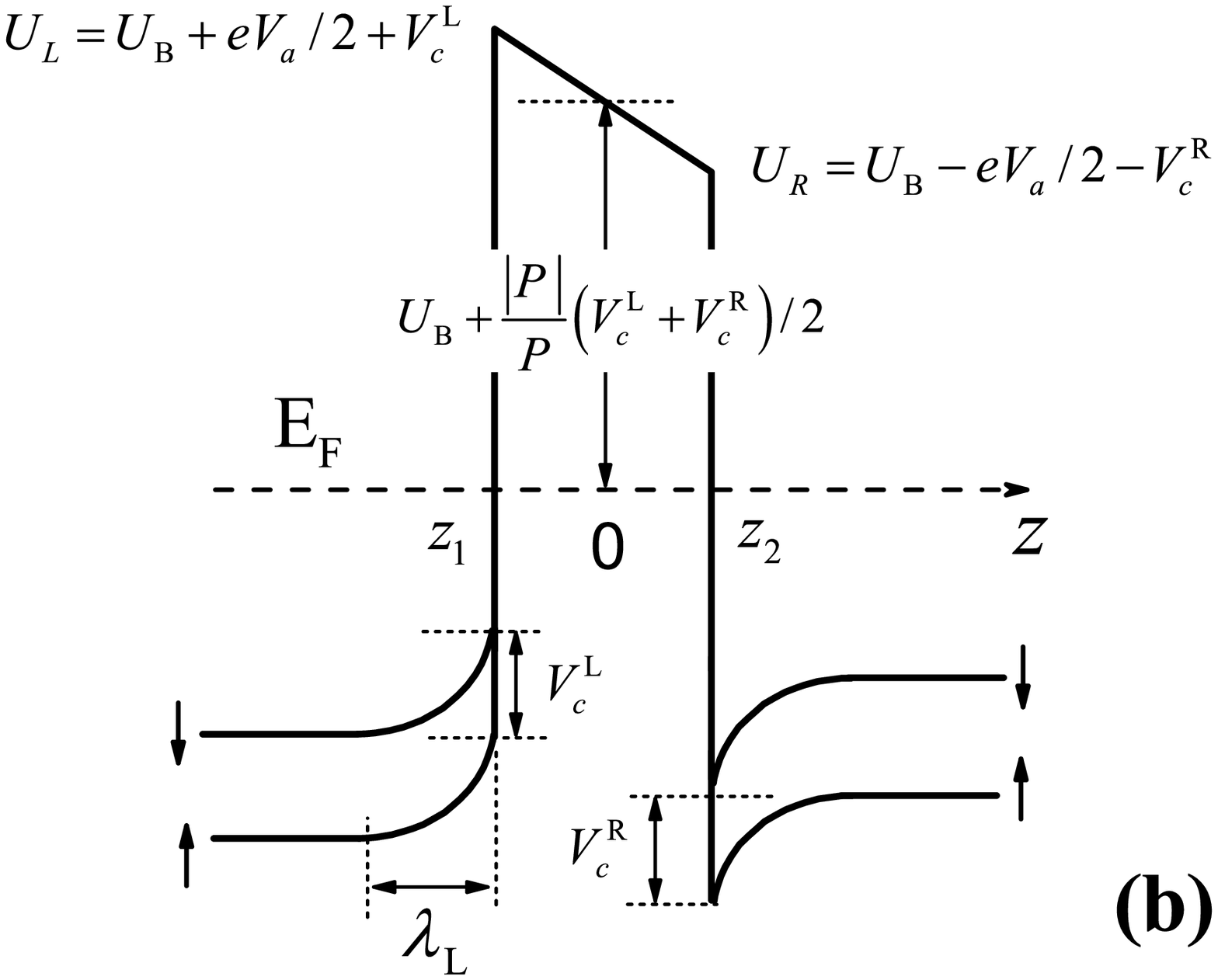}
\end{tabular}
\caption{(a) Schematics of a magnetic tunnel junction comprising two ferromagnets
and a ferroelectric insulator; (b) Potential profile of the junction with positive polarization (screening) and applied voltage $V_a$.\label{fig:fig1}}
\end{figure}

Nonetheless, metal-oxides barriers display a wide variety of functional properties among which ferroelectricity can be used to add a degree of freedom to the system \cite{multiferroics}. In oxide perovskytes, such as BaTiO$_3$, uncompensated charge screening in the material induces a ferroelectric polarization which produces in turn very narrow charge depletion regions at the interfaces, as illustrated in Fig.\,\ref{fig:fig1}. Inserted between non-magnetic metallic electrodes, ferroelectric barriers are expected to result in large tunneling electroresistance effects (TER) \cite{esaki, zhu2, bilc}. The combination of ferroelectricity and ferromagnetism in a same material, referred to as a {\em multiferroic}, allows for the electric manipulation of the magnetization and vice-versa \cite{multiferroics}. However, the difficulty of growing multiferroic thin films limits their exploitation as viable memory elements for now. A promising route is to use {\em synthetic} multiferroics, where a ferroelectric barrier is inserted between two ferromagnets. In this case, the system displays gigantic TER ratios \cite{tsymbal,Gar1,Gar2} resulting from the impact of the ferroelectric polarization on interfacial spin polarization \cite{zhu2,burton,int}. Important progress have been realized recently and ferroelectric junctions have been implemented in solid-state memory devices \cite{chanth}.\par

Whereas the above results are obtained for thick ferroelectric barriers ($>2$ nm), reducing the barrier thickness would enable spin transfer to occur. In the present Letter, we theoretically address the nature of the spin torque in magnetic tunnel junctions comprising such a thin ferroelectric barrier. We demonstrate that the sign and magnitude of the spin transfer torque and magnetoresistance can be controlled by tuning the ferroelectric polarization, opening new avenues for spin-based functional devices.\par

Let us consider a magnetic tunnel junction composed of FM$_{\rm L}$/FEB/FM$_{\rm R}$, where FM$_{\rm L(R)}$ is the semi-infinite left (right) ferromagnetic electrode and FEB is a non-magnetic ferroelectric insulator, such as BaTiO$_3$, PbTiO$_3$, Pb(Zr$_{0.2}$Ti$_{0.8}$)O$_3$, etc. The unit magnetization vectors are ${\bf M}_{\rm L}={\bf{e}}_z$  and ${\bf M}_{\rm R} ={\bf{e}}_x\sin\theta+{\bf{e}}_z\cos\theta$, $\theta$ being the angle between the magnetization directions, see Fig. \ref{fig:fig1}(a).  The junction is treated within a 1-dimensional free-electron two-band Stoner model, following  Refs. \onlinecite{zhang} and \onlinecite{zhu1}. The Hamiltonian of the full system reads ${\hat H}={\hat H}_{\rm L}+{\hat H}_{\rm B}+{\hat H}_{\rm R}$, where

\begin{eqnarray}
\label{1}
\hat{H}_{\rm B}&=&\frac{\hat{\bf p}^2}{2m_{\rm B}}+U_{\rm B}+{\rm E}_{\rm F}-\frac{z}{d}eV_a+\\&&+\left(V_{\rm c}^{\rm L}-\left(\frac{z}{d}+\frac{1}{2}\right)(V_{\rm c}^{\rm R}+V_{\rm c}^{\rm L})\right)\frac{|P|}{P}\nonumber\\
\label{2}
{\hat H}_{\rm L(R)}&=&\frac{\hat{\bf p}^2}{2m}-\frac{\Delta}{2}\hat{\bm \sigma}\cdot{\bf M}_{\rm L(R)} \pm \frac{{eV_a }}{2}  \pm \frac{{\left| P \right|}}{P}eV_c^{{\rm{L}}\left( {\rm{R}} \right)} \times \\ 
&&e^{ - \left( {z \pm \,d/2} \right)/\lambda _{{\rm{L}}\left( {\rm{R}} \right)} }  \left( {1 - \frac{{J\left( {\rho ^ \uparrow   - \rho ^ \downarrow  } \right)}}{{1 + J\rho }}\left( {{\hat{\bm \sigma}} \cdot {\bf{M}}_{{\rm{L}}\left( {\rm{R}} \right)} } \right)} \right),\nonumber
\end{eqnarray}

where 
\begin{eqnarray}
V_{\rm c}^{\rm L(R)}&=&\frac{d\lambda_{\rm L(R)}|P|}{\epsilon_0d+\epsilon_{\rm f}(\lambda_{\rm L}+\lambda_{\rm R})},\\
\label{3}
\lambda_i&=&\frac{e^2}{\epsilon_{\rm 0}}\frac{\rho_{i}+4J_{i}\rho_{i}^\downarrow\rho_{i}^\uparrow}{1+J_{i}\rho_{i}}.
\end{eqnarray} 
$\lambda_{i}$ is the screening length of the $i$-th electrode, $\rho^\sigma$ is the spin-dependent density of states with $\rho=\rho^\uparrow+\rho^\downarrow$, $d$ is the thickness of the ferroelectric barrier, $\hat{\bm \sigma}=\left( {\hat \sigma _x ,\hat \sigma _y ,\hat \sigma _z } \right)$ is the vector of the Pauli matrices, $\epsilon_{\rm 0},\epsilon_{\rm f}$ are the dielectric static permittivity of the vacuum and ferroelectric barrier, respectively. $P$ is the ferroelectric polarization, $\Delta=J m_{\rm sd}$ is the exchange energy, ${\rm{m}}_{{\rm{sd}}} {\rm{ = }}\left( {{\rm{n}}^ \uparrow   - {\rm{n}}^ \downarrow  } \right)$ is the {\em equilibrium} spin density, ${\rm{n}}^{ \uparrow {\rm{,}} \downarrow }  = \frac{1}{{6\pi ^2 }}\left[ {\frac{{2m_e }}{{\hbar ^2 }}\left( {{\rm{E_F}}  \pm \Delta /2} \right)} \right]^{3/2}$, $J$ is Stoner exchange parameter, the hat $\hat{}$ denotes a 2$\times$2 matrix in spin space.\par

The eigenstates of the full system ${\hat H}$ are obtained by standard wave matching procedure and the resulting spin-dependent wave functions $\hat{\phi}_{\bf k}$ are expressed in terms of a combination of Airy (barrier) and Bessel (ferromagnets) functions. In this picture, the charge and spin currents are defined
\begin{eqnarray}
&&j_e=-\frac{e\hbar}{m}\Im\sum_{i=\rm{L,R}}\int\frac{d^3{\bf k}}{(2\pi)^3}\hat{\psi}_{\bf k}^\dagger\hat{\bm \nabla}\hat{\psi}_{\bf k}f_{i}({\bf k}),\\
&&{\cal J}=\frac{\hbar^2}{2m}\Im\sum_{i=\rm{L,R}}\int\frac{d^3{\bf k}}{(2\pi)^3}\hat{\psi}_{\bf k}^\dagger[\hat{\bm \sigma}\otimes\hat{\bm \nabla}]\hat{\psi}_{\bf k}f_{i}({\bf k}),\label{eq:j}
\end{eqnarray}
$\Im$ referring to the imaginary part and $f_{i}({\bf k})$ being the Fermi-Dirac distribution function of the $i$-th electrode. The spin transfer torque is defined as \cite{slonc96} ${\bf T}=-\int_\Omega{\bm\nabla }\cdot{\cal J}d\Omega$, where $\Omega$ is the volume of the ferromagnet. In the case of semi-infinite electrodes, it reduces to the interfacial tunneling spin current: ${\bf T}=T_{||}\,{\bf M}_{\rm R}\times({\bf{e}}_z\times{\bf M}_{\rm R})+T_\bot{\bf M}_{\rm R}\times{\bf{e}}_z$, where $T_{||}  = {\cal J}_{\rm int}^{x}$ and $T_ \bot   = {\cal J}_{\rm int}^{y} $. For the numerical simulations, we choose the parameters describing Fe/BaTiO$_3$/Fe symmetric MTJs  (at low temperatures): the effective Fermi energy is E$_{\rm F}$=2.62 eV with an exchange splitting $\Delta$=3.86 eV corresponding to the majority (minority) Fermi wave vector $k^{\uparrow}_{\rm F}$=1.09 \AA$^{-1}$ ($k_{\rm F}^\downarrow$=0.4 \AA$^{-1}$) and $J =1.07$ eV. The barrier dielectric permittivity is $\epsilon_{\rm f}$=90$\epsilon_{\rm 0}$ and its thickness is $d$ =1 nm. The barrier height and electron effective mass are taken as $U_{\rm B}$=0.6 eV and  $m_{\rm B}/m_{\rm e}$=0.28. These parameters, rather conservative, are extracted from {ab-initio} calculations and correspond to an effective decay wave vector of $q\approx$2 nm$^{-1}$ \cite{burton}. The screening lengths were derived according to Eq.(\ref{3}), $\lambda_{\rm L(R)}$=0.8 \AA. The typical ferroelectric polarizations (FEP) highlighted in the literature are in the range of 10 to 75 ${\rm{\mu C/cm}}^{\rm{2}}$ \cite{Gar2,Petraru}. The value of the FEP strongly depends on the ferroelectric layer thickness and interfacial structure \cite{Petraru,burton,int}. Finally, in the present work, we only focus on the case of a {\em symmetric} tunnel junction composed of similar ferromagnetic electrodes.\par

\begin{figure}
\includegraphics[width=6.2cm]{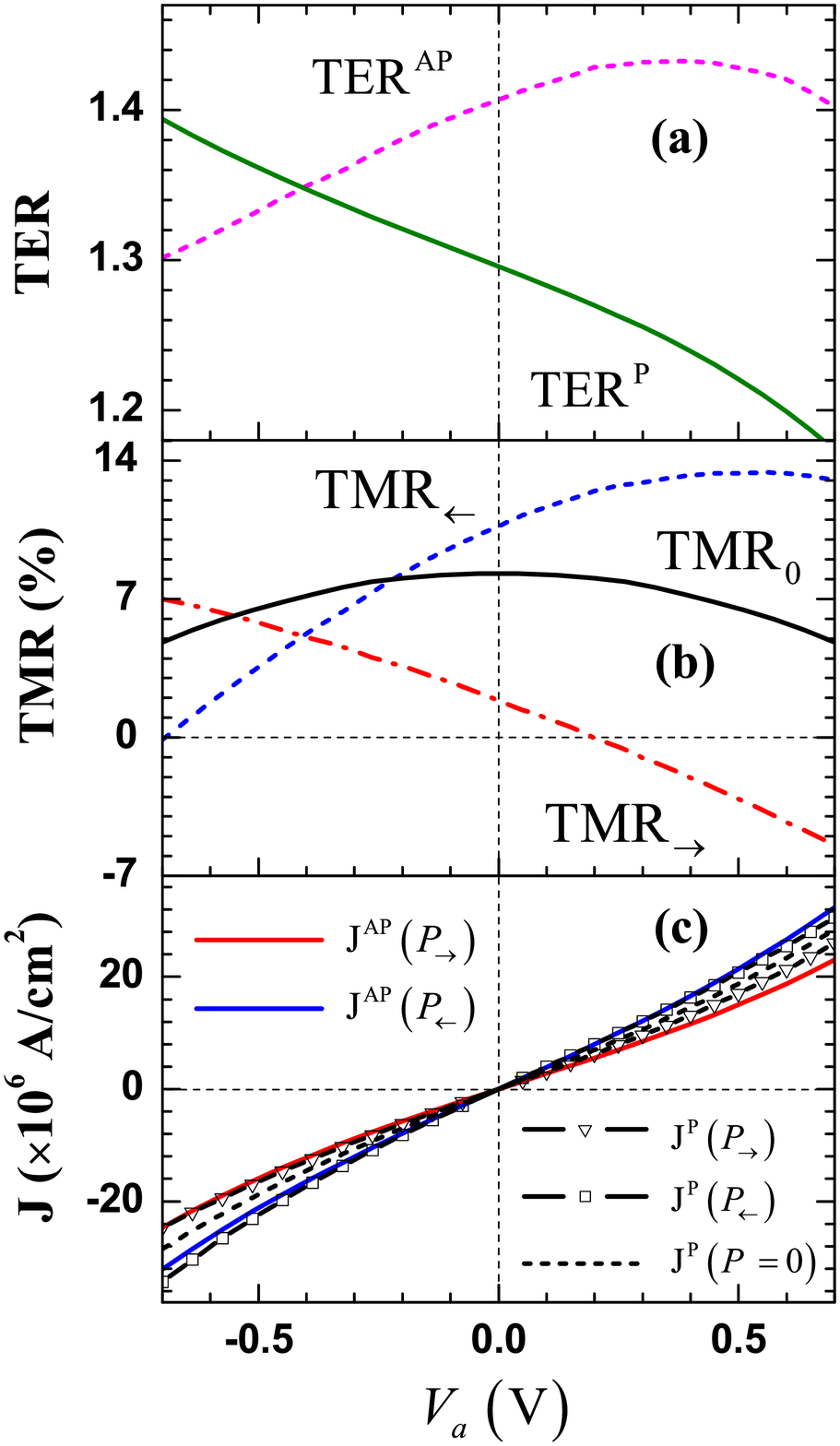}
\caption{(Color online) Voltage dependence of (a) TER for P (solid line) and AP (dashed line) configurations and (b) TMR for zero (solid line), positive (dotted-dashed line) and negative FEP (dotted line). (c) Corresponding current-voltage characteristics for different polarization and magnetic configurations, as indicated on the figure. The parameters are the same as in the text, with $P$=75 $\mu$C/cm$^{2}$, and $V_c^{\rm{L,R}}$=0.44 eV. \label{fig:fig2}}
\end{figure}

The interplay between tunnel electroresistance and spin transport has been predicted theoretically \cite{zhu1,burton,zhu2} and demonstrated experimentally \cite{chanth, Gar1,Gar2}, allowing for the ferroelectric control of the magnetoresistance and the magnetic control of the electroresistance. However, the influence of the ferroelectric polarization on the bias dependence of the tunneling magnetoresistance remains unexplored. In the remainder of this work, we define the TMR and TER as
\begin{eqnarray}
&&\rm{TMR}_{ \mathbin{\lower.3ex\hbox{$\buildrel\textstyle\rightarrow\over
{\smash{\leftarrow}\vphantom{_{\vbox to.5ex{\vss}}}}$}}} =\frac{{J^{\rm{P}} (P_\leftrightarrow)-J^{{\rm{AP}}} (P_\leftrightarrow) }}{{J^{{\rm{AP}}}  (P_\leftrightarrow)}}\times 100\,\% ,\\
&&{\rm{TER}}^{{\rm{P}}\left( {{\rm{AP}}} \right)}  = \frac{{{\rm{J}}^{{\rm{P}}\left( {{\rm{AP}}} \right)} \left( {P_ \to} \right)}}{{{\rm{J}}^{{\rm{P}}\left( {{\rm{AP}}} \right)} \left( {P_ \leftarrow  } \right)}}.
\end{eqnarray}
$ {\rm{J}}^{{\rm{P}}\left( {{\rm{AP}}} \right)} \left( {P_ \leftrightarrow  } \right)$ denotes the tunneling charge current in parallel (P) and antiparallel (AP) configurations, and $P_{\leftrightarrow}$ refers to the leftward (negative) or rightward (positive) ferroelectric polarization of the barrier. The bias dependence of the tunnel magneto- and electroresistance as well as the current-voltage characteristics are shown in Fig.\,\ref{fig:fig2}(a)-(c). The TER evaluated for this symmetric MTJ is rather small : TER$\approx$1.3 (30\%) for P and TER$\approx$1.4 (40\%) for AP magnetic configurations [Fig.\,\ref{fig:fig2}(a)]. Switching the magnetization configuration results in a strong asymmetric alteration of the bias dependence of the electroresistance effect. The most striking feature is obtained when analyzing the influence of the ferroelectric polarization on the TMR. In the absence of polarization [Fig.\,\ref{fig:fig2}(b), black curve], the TMR is symmetric as expected in the case of a symmetric MTJ \cite{review_mtj}. We note that the bias-induced TMR reversal occurs at large bias voltages ($V_a>$1 V, not shown). However, in the presence of ferroelectric polarization in the barrier, the potential profile becomes asymmetric and the bias dependence of the TMR is heavily distorted, yielding TMR sign reversal at small voltages. Asymmetric tunnel junctions are expected to generate much larger TER effects than symmetric ones \cite{zhu1} and therefore, one can expect much stronger distortion of the bias dependence of the TMR, depending on the FEP direction.

\begin{figure}
\includegraphics[width=6.2cm]{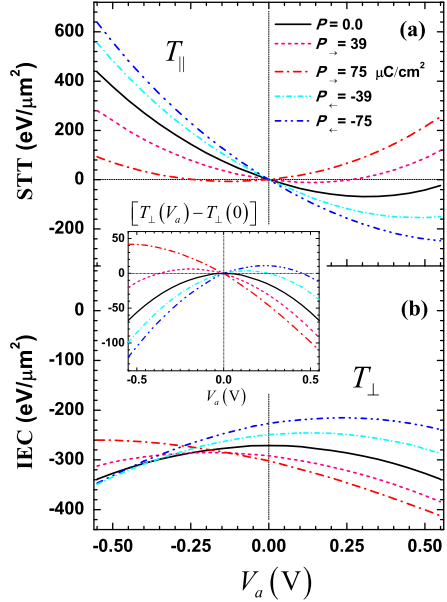}
\caption{(Color online) Voltage dependence of the (a) in-plane and (b) out-of-plane torques exerted on the right layer for different values of FEP, as indicated on the figure.  Inset: non-equilibrium out-of-plane torque component. The parameters are the same as in Fig. \ref{fig:fig2} and $\theta  = \pi /2$. \label{fig:fig3}}
\end{figure}

\begin{figure}
\includegraphics[width=6cm]{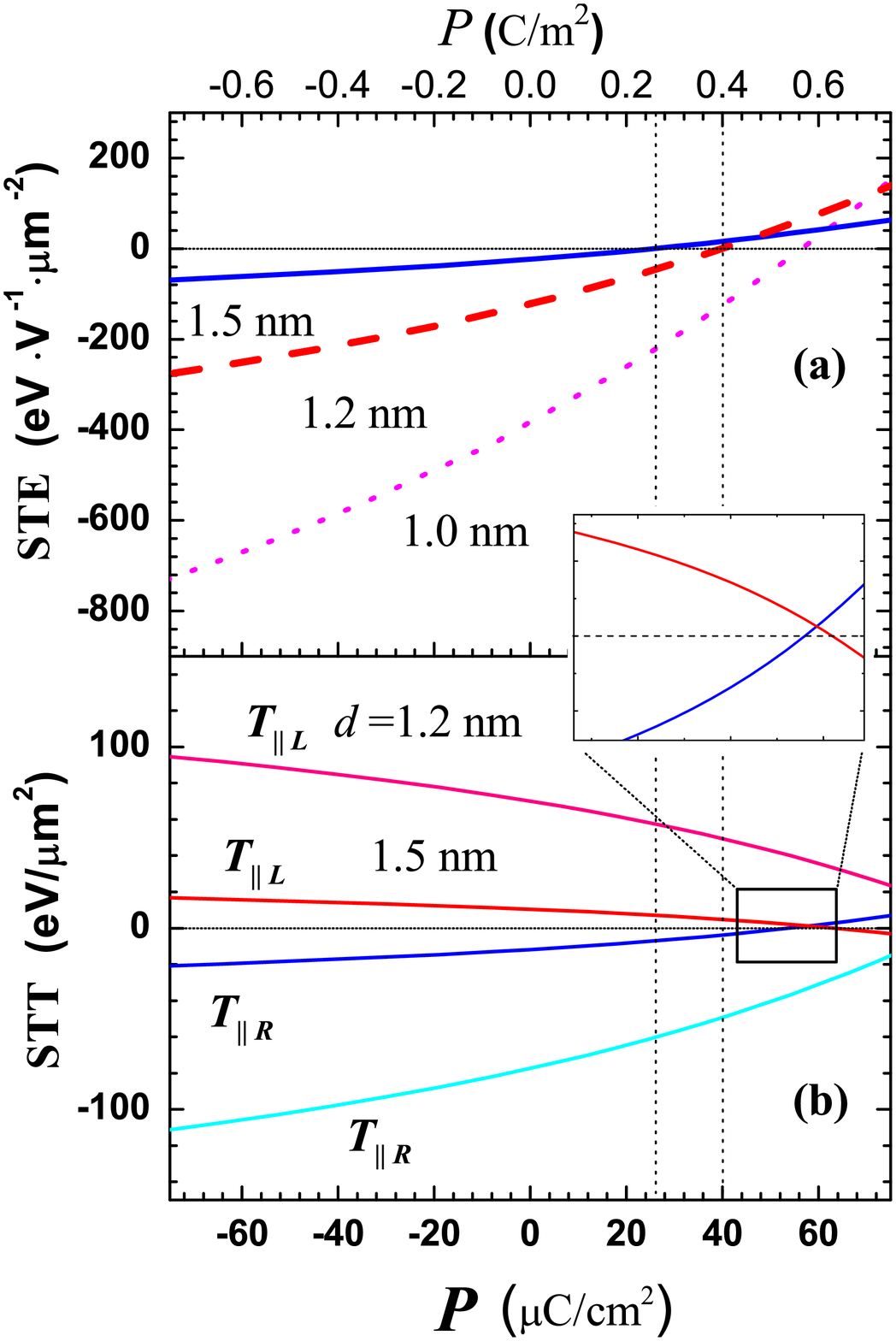}
\caption{(Color online) (a) Low bias spin torque efficiency as a function of the FEP for barrier thicknesses ranging from 1 to 1.5 nm; (b) Contribution from rightward ($T_{||\rm L}$) and leftward ($T_{||\rm R}$) electrons to the in-plane torque exerted on the right layer as a function of the FEP for 1.2 and 1.5 nm-thick barriers. Inset: zoom of the curves crossing point for the 1.5 nm barrier case. The vertical dotted lines connect the STE reversal with the compensation of point of leftward and rightward electrons. The parameters are the same as in Fig. \ref{fig:fig3}. \label{fig:fig5}}
\end{figure}

Let us now consider the spin transfer torque exerted on the right layer. Figure \ref{fig:fig3}(a) and \ref{fig:fig3}(b) display the bias dependence of the in-plane $T_ {||}  \left( {V_a } \right)$ and  out-of-plane $T_\bot \left( {V_a } \right)$ components, respectively. In the absence of ferroelectricity (solid lines), the torque has the form $T_ {||}=a_1V_a+a_2V_a^2$ and $T_{\bot}=b_0+b_2V_a^2$, as expected in the case of a symmetric tunnel junction \cite{Theo,asym, spindiode}. In the presence of ferroelectric polarization, the bias dependence of both components is strongly rectified. The interlayer exchange coupling constant, defined as $ {T_ \bot  \left( {V_a } \right) } $ \cite{slonc96,Theo}, decreases in absolute value for negative FEP and increases for positive one, in agreement with Ref. \onlinecite{zhu2Cond}. Most importantly, we find that the spin torque efficiency (or {\em torkance} \cite{spindiode}), defined as ${\rm{STE}} = \partial T_{||} /\partial V_a \left( {V_a  \to 0} \right)$, is dramatically affected by the presence of ferroelectric polarization (see Fig. \ref{fig:fig5}). The non-equilibrium interlayer exchange coupling (or net spin torque) defined as $\left[ {T_ \bot  \left( {V_a } \right) - T_ \bot  \left( 0 \right)} \right]$  acquires a positive (negative) efficiency for negative (positive) ferroelectric polarization for small biases (inset of Fig. \ref{fig:fig3}). The in-plane torque efficiency is also rectified and can be even reversed at large FEP, as displayed in Fig. \ref{fig:fig3}(a).

These results demonstrate the control of the spin torque efficiency by the ferroelectric polarization. In Fig. \ref{fig:fig5}(a) the FEP-dependence of the in-plane torque efficiency is represented for different barrier thicknesses. The slope of the spin torque efficiency as a function of the FEP decreases at larger thickness, indicating that the ferroelectric control of the torkance is more effective in thin barriers. In other words, the difference in spin torque magnitude when switching the FEP is larger when decreasing the barrier thickness. On the other hand, the threshold FEP required to reverse the sign of the in-plane torque decreases with the barrier thickness.\par

These two behaviors can be interpreted by analyzing the asymmetric spin transport in the barrier. In the non-equilibrium transport formalism, the total spin torque exerted on the right ferromagnet arises from a balance between the torque contributions of the electrons originating from the left reservoir and those originating from the right reservoir [see Eq. (\ref{eq:j})]. In the presence of ferroelectricity, the barrier becomes asymmetric and the tunneling rate of leftward and rightward electrons is no more equal, resulting in the TER \cite{zhu1,zhu2}. The contributions of the electrons originating from the left and right reservoirs to the in-plane torque as a function of the FEP are displayed in Fig. \ref{fig:fig5}(b), for different barrier thicknesses. The slope of the spin torque as a function of the FEP is larger for thinner barriers, due to the larger amount of tunneling electrons experiencing the asymmetric tunneling. On the other hand, the TER effect is more effective for thicker barriers and the asymmetry in tunneling probability between leftward and rightward electrons is larger. This increased tunneling electroresistance results in a reversal of the sign of the in-plane torque at positive FEP.\par

The ferroelectric control of the bias dependence of both in-plane and out-of-plane components of the spin torque has important impacts in terms of current-driven magnetization switching and dynamics. It is usually accepted that in-plane torque drives the magnetization switching and self-sustained precessions \cite{Brataas,Ralph}. Therefore, a smart design of the ferroelectric junction enables the control of the dynamical properties by tuning both the injected current and ferroelectric polarization. Another interesting aspect is the control of the out-of-plane torque. In a recent publication, Oh et al. \cite{Oh} demonstrated the important role of the out-of-plane component on current-driven magnetic instabilities (tagged {\em switching back phenomenon}) in MTJs. In light of the present study, controlling this effect through ferroelectricity seems a viable route to reduce these instabilities. Since the ferroelectric polarization is usually controlled through voltage pulses, spin torque and FEP switching can be combined to obtain original device operation modes. \par

Finally, two important aspects are to be considered regarding the applicability of the present theory to experimental situations. First, the TER obtained numerically with the present model (two band Stoner model in a symmetric junction) is small compared to the experimental observations: about 40\% in the present work compared to 10,000\% \cite{chanth} and 75,000\% \cite{Gar1} in thick BaTiO$_3$. This indicates that the asymmetry expected experimentally is actually much larger than the one derived in the present work, which implies that the ferroelectric control of the spin torque efficiency is expected to be much larger. Second, whereas the voltage-control of tunneling electroresistance has been recently realized in thick BaTiO$_3$-based MTJs \cite{Gar1,chanth}, the realization of spin transfer torque in thin BaTiO$_3$ junctions presents the advantage of generating unique current-driven dynamical states that are absent in thick ferroelectric junctions. The combination of ferroelectricity with spin transfer torque is thus expected to produce a wealth of original states of major interest for technologists and researchers.

We thank Prof. M. Zhuravlev and Dr. N. Useinov for useful conversations and support, and Prof. E.Y. Tsymbal for his critical review of the manuscript and his constructive comments.

\end{document}